\documentclass[pdflatex,sn-mathphys-num]{sn-jnl}

\usepackage{graphicx}%
\usepackage{multirow}%
\usepackage{amsmath,amssymb,amsfonts}%
\usepackage{amsthm}%
\usepackage{mathrsfs}%
\usepackage[title]{appendix}%
\usepackage{xcolor}%
\usepackage{textcomp}%
\usepackage{manyfoot}%
\usepackage{booktabs}%
\usepackage{algorithm}%
\usepackage{algorithmicx}%
\usepackage{algpseudocode}%
\usepackage{listings}%

\theoremstyle{thmstyleone}%
\theoremstyle{thmstyletwo}%

\theoremstyle{thmstylethree}%

\usepackage{caption}
\usepackage{subcaption}
\usepackage{xcolor}
\usepackage{booktabs}
\usepackage{multirow}
\usepackage{graphicx}
\usepackage[pagewise]{lineno}	
\usepackage{soul}

\begin{document}

\title[Exploring the Effects of Color Reconnection on Net-Charge Fluctuations in $pp$ Collisions at $\sqrt{s}=2.76 \text{ TeV}$ using PYTHIA Monash]{Exploring the Effects of Color Reconnection on Net-Charge Fluctuations in $pp$ Collisions at $\sqrt{s}=2.76 \text{ TeV}$ using PYTHIA Monash}


\author[1]{\fnm{Dibakar} \sur{Dhar}}

\author[1]{\fnm{Tumpa} \sur{Biswas}}

\author[2]{\fnm{Zubayer} \sur{Ahammed}}

\author*[1]{\fnm{Prabir} \sur{Kumar Haldar}}\email{prabirkrhaldar@gmail.com}

\affil*[1]{\orgdiv{Department of Physics}, \orgname{Cooch Behar Panchanan Barma University}, \orgaddress{\street{Vivekananda Street}, \city{Cooch Behar}, \postcode{736101}, \state{West Bengal}, \country{India}}}

\affil[2]{\orgname{Variable Energy Cyclotron Centre (VECC)}, \orgaddress{\street{HBNI, 1/AF Bidhannagar}, \city{Kolkata}, \postcode{700064}, \state{West Bengal}, \country{India}}}



\abstract{This work explores the effects of various color reconnection (CR) configurations on the net charge fluctuations of charged particles generated in $pp$ collisions at a center-of-mass energy of $\sqrt{s}$ = 2.76 TeV, simulated with the PYTHIA Monash using the variable $\nu_{(+,-,\text{dyn})}$. The results are compared with experimental data from ALICE, revealing that the current PYTHIA simulation settings overestimate the net charge fluctuations relative to the experimental observations. Furthermore, $\nu_{(+,-,\text{dyn})}$ exhibits a clear dependence on the MPICR range parameter, where increasing the range leads to more negative $\nu_{(+,-,\text{dyn})}$ values, suggesting the geometric extent of CR directly modulates the strength of local charge correlation. To ensure these effects are not a trivial consequence of particle density, a comparison with a retuned NoCR baseline at fixed multiplicity was performed. This analysis confirms that CR induces a genuine dynamical shift in fluctuations that persists independently of the average charged-particle density. The study also shows that the net charge and net pion channels display the strongest suppression in $\langle N_{\text{ch}} \rangle \nu_{(+-,\text{dyn})}$ compared to the net kaon and net proton channels. Overall, the study offers a useful benchmark against experimental results and sheds light on the underlying dynamics of the PYTHIA Monash.}

\keywords{Net-charge Fluctuation, QGP, QCD, PYTHIA Monash}



\maketitle

\section{Introduction}

For decades, high-energy physics experiments such as the Super
Proton Synchrotron (SPS) \cite{1,2}, the Relativistic Heavy-Ion Collider (RHIC) \cite{3,4,5,6,7,8,9}, and the Large Hadron Collider (LHC) \cite{10,11,12,13,cms} have been working with a primary objective to explore as many signals as possible toward characterizing the properties of the quark-gluon plasma (QGP), which is a deconfined state of quarks and gluons produced in high-energy heavy-ion collisions. Event-by-event fluctuations are one of the most significant tools for characterizing the thermodynamic properties of this system \cite{14,15,16,17,s0,s1,s2,s3,s4,s5}. Specifically, fluctuations in conserved quantities within a limited phase space, such as the net charge of the system, have long been studied as potential indicators of QGP formation and phase transitions \cite{18,19,20,21,22,23}. However, these early estimates were rather preliminary, as they did not include the effects of resonance decays, local charge conservation, or kinematic acceptance cuts. Recent work has addressed these limitations by developing formalisms that incorporate these crucial correlation effects \cite{29,d0,d1,d2,d3,d4,d5,d6,d7,d8}. In particular, J. Parra \textit{et al.} \cite{d9} showed that while the hadron gas scenario requires a very short range of local charge conservation to describe ALICE data, the QGP scenario agrees with the data across a wide range of parameters, with a Bayesian analysis yielding moderate evidence for QGP freeze-out of charge fluctuations.

The fluctuations in the net charge depend on the sum of the squared charges of the particles in the system. Under certain assumptions (e.g., a quench scenario where hadronization has no influence on fluctuations), net-charge fluctuations in the HRG phase are predicted to be significantly greater than in a QGP phase \cite{16}. If the initial QGP phase is strongly dominated by gluons, the fluctuation per unit of entropy may decrease further as the hadronization of gluons increases entropy \cite{17}. So, the net-charge fluctuations are significantly influenced by the phase from which they arise.
However, these fluctuations can also be affected by event-by-event fluctuations in the number of particle-production sources (such as the number of multiparton interactions per event) \cite{d4,d5} and by the subsequent final-state interactions \cite{d6,d7,d8} that the produced particles may undergo. To address this, one may consider the fluctuations in the ratio, $R = N_+/N_-$, where $N_-$ and $N_+$ are the numbers of the negative and positive particles, respectively, within a confined pseudorapidity and transverse momentum range. A commonly used tool for assessing the strength of these fluctuations is the $D$ measure, which can be defined as \cite{14,15,16}
\begin{equation}
\label{eqn1}
D = \langle N_{ch} \rangle \langle \delta R^2 \rangle \approx \frac{4 \langle \delta Q^2 \rangle}{\langle N_{ch} \rangle},
\end{equation}
where $Q = N_+ - N_-$ is the net charge, $\langle \delta Q^2 \rangle$ is the variance of $Q$, and $\langle N_{ch} \rangle = \langle N_+ + N_- \rangle$ is the event average number of charged particles. The D measure has been estimated based on various theoretical frameworks, and it has been observed that this D measure is approximately four times greater for the HRG phase compared to the QGP phase \cite{16}. Another notable observable for studying net charge fluctuations is $ \nu_{(+-,dyn)} $, which represents the relative correlation of particle pairs and can be defined as  \cite{22,23,24,25,26,27,28}

\begin{equation}
\label{eqn2}
\nu_{(+-,\text{dyn})} = \frac{\langle N_+(N_+-1)\rangle}{\langle N_+\rangle^2} + \frac{\langle N_-(N_--1)\rangle}{\langle N_-\rangle^2} - 2\frac{\langle N_-N_+\rangle}{\langle N_-\rangle\langle N_+\rangle}.
\end{equation}

A negative value of $\nu_{(+-,dyn)}$ indicates that the correlations primarily arise from pairs of opposite charges. In contrast, a positive value suggests that the correlations are mainly due to pairs of the same charge. Also, it is important to note that the measured ratio ($\nu_{+-,dyn}$ variable) is independent of single particle detector inefficiency \cite{23}. The $\nu_{(+-,dyn)}$ and the D measure are related to each other by \cite{15}

\begin{equation}
\langle N_{ch} \rangle \nu_{(+-,dyn)} \approx D-4.
\end{equation}

The values of $\nu_{(+-, dyn)}$ must be corrected for global charge conservation and finite acceptance effects \cite{22,29}. If all charges were accepted, global charge conservation would lead to vanishing fluctuations, yielding a minimum value of $\nu_{(+-, dyn)}$ to be $-4/\langle N_{\text{total}} \rangle$, where $\langle N_{\text{total}} \rangle$ is the average total number of charged particles produced over the full phase space. The corrected $\nu_{(+-, dyn)}$ is then given by

\begin{equation}
\nu_{(+-, dyn)}^{\rm corr} = \nu_{(+-, dyn)} + \frac{4}{\langle N_{\text{total}} \rangle}.
\label{corrected}
\end{equation}

\textcolor{black}{It is important to note that this correction, as derived by Pruneau \textit{et al.} \cite{22}, is valid under the assumptions that the charge acceptance is symmetric about mid-rapidity, that global charge conservation is the sole source of the finite-acceptance bias, and that the total charged-particle multiplicity $\langle N_{\rm total} \rangle$ is evaluated over the full $4\pi$ acceptance. Within these assumptions, the correction is additive and independent of the experimental acceptance used to measure the net-charge fluctuations. This is the same correction applied by the ALICE Collaboration in their analysis of Pb-Pb collisions \cite{29}.}

In a study from the STAR Collaboration \cite{23} on net-charge fluctuations in $Au+Au$ collisions at different center-of-mass energies from 19.6 to 200 GeV in a symmetric range about mid-rapidity and with relative particle pseudorapidity $<$ 1.0, the results are consistent with the HRG theoretical predictions. Similarly, from the study of the ALICE Collaboration \cite{29} in $Pb-Pb$ collisions at $\sqrt{s_{\rm NN}}$ = 2.76 TeV in terms of  $\nu_{(+-,dyn)}$ in symmetric ranges about mid-rapidity with $\Delta \eta$ = 1.0 and 1.6 ranges, consistent with the HRG and close to the QGP limits, respectively. Also, S. Paul, \textit{et al.} \cite{28} studied charged particle fluctuations for $Pb-Pb$ collisions at $\sqrt{s_{\rm NN}}=5.5$ TeV using the AMPT model, and their results are also close to the QGP limits. While these studies cover heavy-ion collision systems, a systematic investigation of net-charge fluctuations using this approach in small systems is limited. So, in this analysis, we have studied the charge particle fluctuation in $pp$ collisions at $\sqrt{s} =$2.76 TeV using PYTHIA v8.3 \cite{30}. We focus on smaller systems, such as $pp$ collisions, because the traditional view held that QGP formation was exclusive to heavy-ion collisions. Small systems, with fewer available partons, were thus treated as the baseline reference to help us understand medium formation in nuclear collisions. However, several recent studies have suggested the possible formation of QGP droplets in small collision systems, particularly in high-multiplicity $pp$ collisions, based on observed heavy-ion-like signatures in these events \cite{31,32,33}. Also, it is worthwhile to mention that here we are studying the effects of Color Reconnection (CR) on the net-charge fluctuation of $pp$ collisions. CR is a mechanism that describes how the fundamental forces of Quantum Chromodynamics (QCD) rearrange the color charges of quarks and gluons before they undergo hadronization \cite{34}. 

\section{Data Description}

The PYTHIA \cite{30,40.2} Monte Carlo event generator is a highly versatile tool based on perturbative quantum chromodynamics (QCD) for simulating high-energy particle collisions. It utilizes a factorized perturbative expansion to model hard parton-parton interactions, complemented by parton shower evolution and comprehensive approaches to hadronization (using the Lund string model) and multiparton interactions. PYTHIA has been widely employed to analyze $pp$ collisions and has demonstrated effectiveness in reproducing various experimental results from collisions at LHC energies. Over the years, different parameters of the existing PYTHIA model have been refined to enhance its accuracy in data description. The PYTHIA Monash tune \cite{39,40,40.1} (i.e., Monash 2013) represents a refined set of parameter values optimized to achieve a more precise description of LHC experimental data. Further, in PYTHIA, color reconnection (CR), a string fragmentation model, has been implemented where the final partons are considered to be color connected in such a way that the total string length becomes as short as possible, and in this work, we have considered different types of CR conditions as follows:

(a) \textbf{PYTHIA NoCR:} PYTHIA NoCR stands for PYTHIA Monash without considering CR = Off, i.e., no color reconnection. Color strings are connected strictly according to the Leading Color (LC) flow from the hard process and parton showers \cite{39}.

(b) \textbf{PYTHIA MPICR:} This stands for PYTHIA Monash with multiparton interactions (MPI) based CR. It is the default CR condition for PYTHIA v8.3. Also, it has an extra parameter ``range" and the higher the value of this ``range" is, the more reconnections can occur. For values above unity, the reconnection rate tends to saturate, since then most systems are already connected. The default value of the ``range" parameter is 1.8. However, to systematically investigate
its influence on net-charge fluctuations, the parameter has been varied over the interval 0.5 -- 5.0 \cite{40}.

(c) \textbf{PYTHIA QCDCR:} This stands for PYTHIA Monash with QCD-based CR. It is based on reconnection based on SU(3) color rules and minimizing the total ``$\lambda$ measure" (an invariant string length proxy). Allows for string junctions to be formed, improving $\Lambda/K$ ratios \cite{41}.

(d) \textbf{PYTHIA GMCR:} This stands for PYTHIA Monash with Gluon-Move CR. It is based on advanced QCD Topology. 
It allows gluons to ``move" and re-link between strings to minimize the $\mathbf{\lambda}$ measure, focusing on local minimization \cite{39}.

It is very important to note that for every setup, PYTHIA Monash is tuned, and the ``range" parameter is only available for PYTHIA MPICR, so for other cases ``range" is not considered.

\begin{figure}[!ht]
\begin{center}
\includegraphics[scale=0.52]{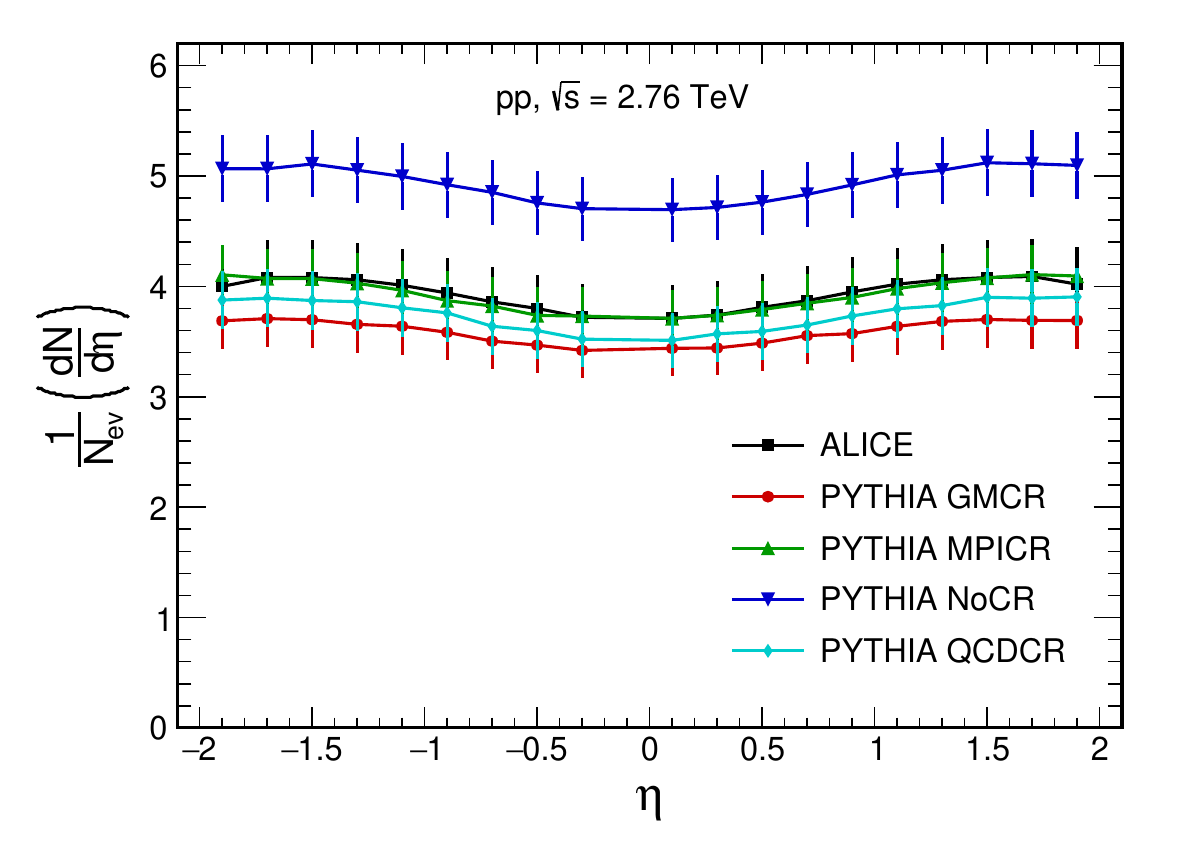} 
\caption{Pseudorapidity distribution of primary charged particles for PYTHIA under various CR tunes, compared with ALICE experimental data \cite{32} in minimum-bias $pp$ collisions at $\sqrt{s} = 2.76$ TeV.}
\label{fig:0}
\end{center}
\end{figure}

\begin{figure*}[ht!]
    \centering
    
    \begin{minipage}{0.48\textwidth}
        \centering
        
        \begin{subfigure}[b]{\linewidth}
            \centering
            \includegraphics[width=\linewidth]{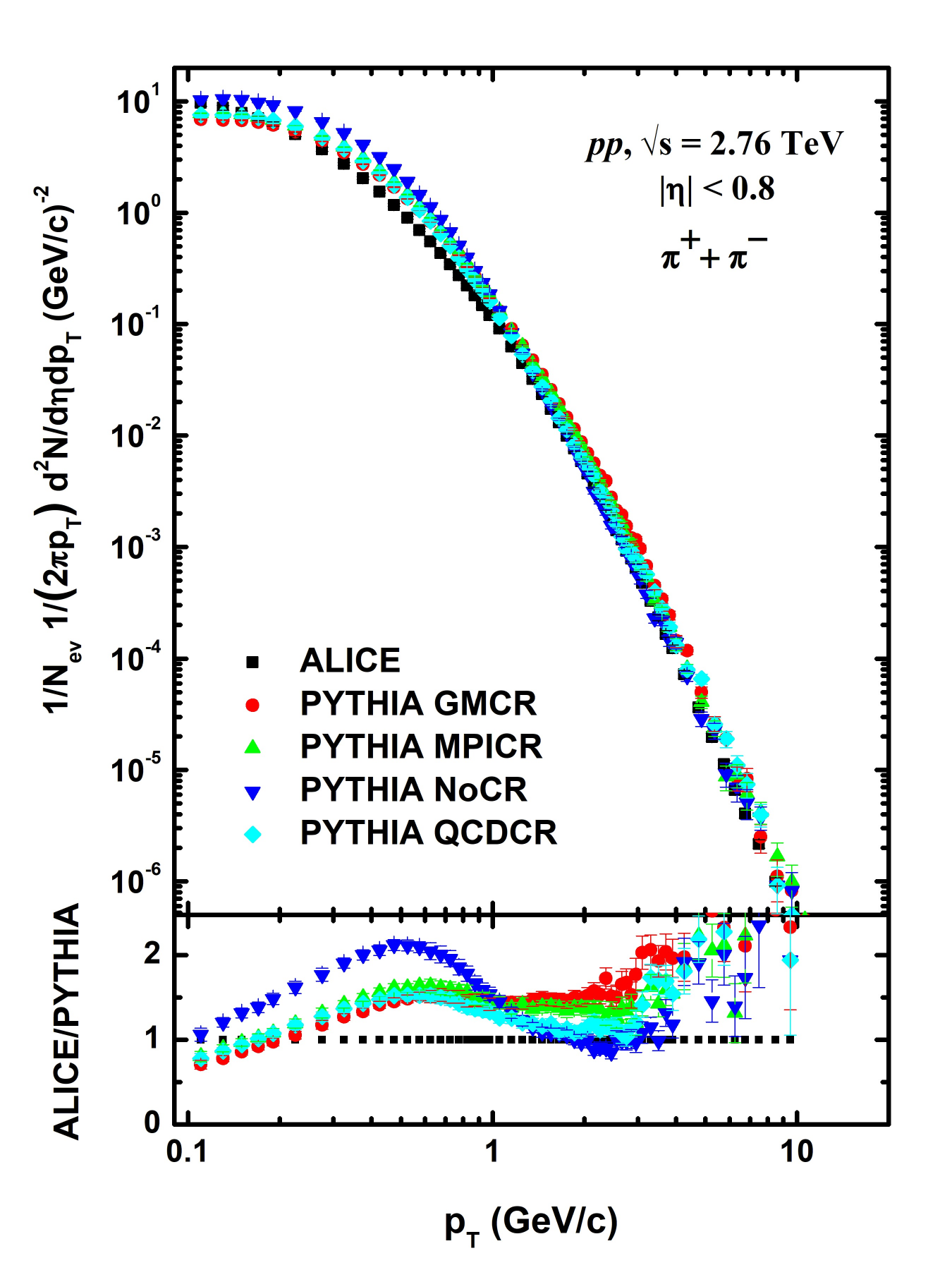}
            \caption{}
            \label{fig:1a}
        \end{subfigure}
        
        \vspace{1em}
        
        \begin{subfigure}[b]{\linewidth}
            \centering
            \includegraphics[width=\linewidth]{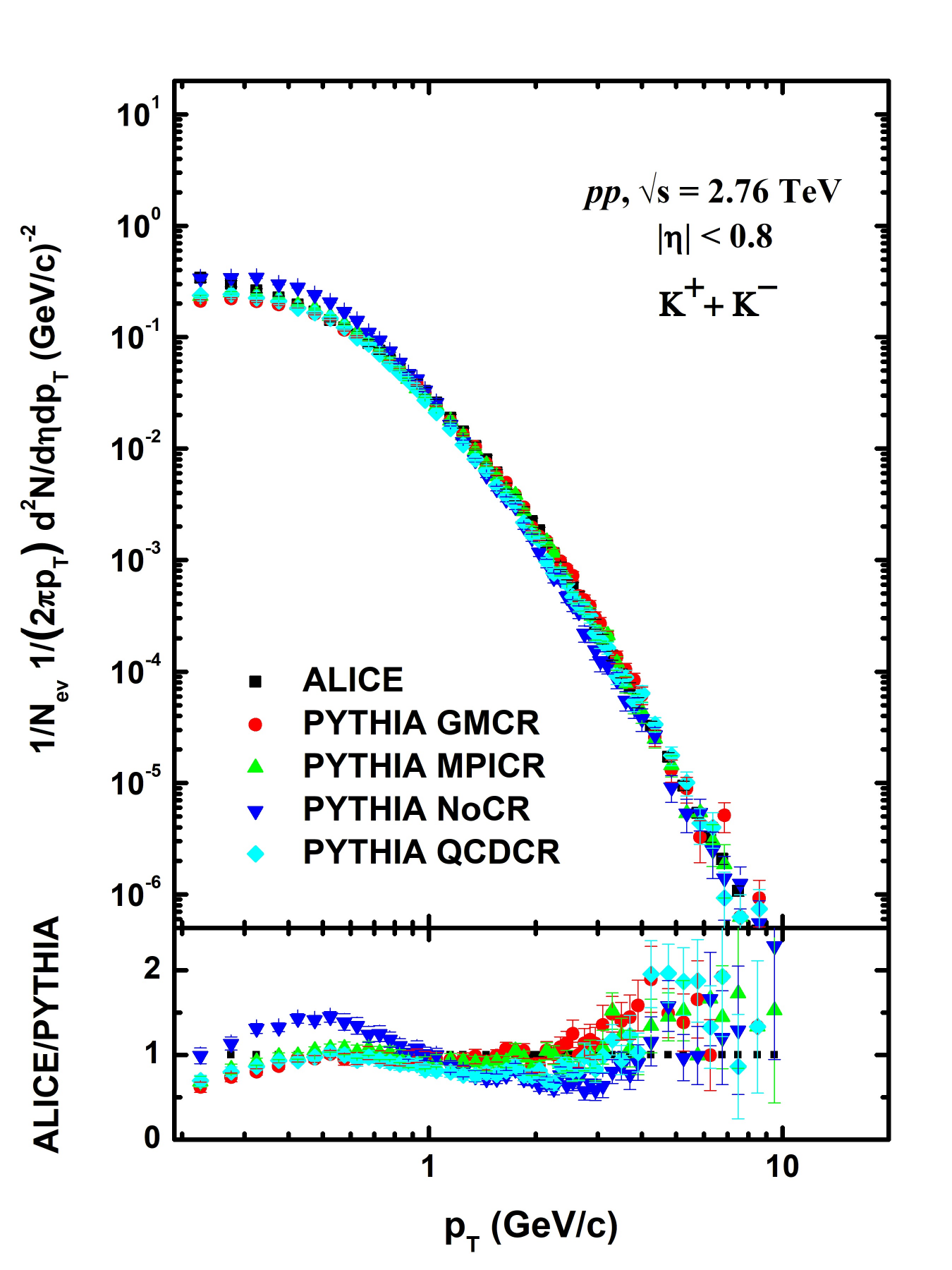}
            \caption{}
            \label{fig:1b}
        \end{subfigure}
        
    \end{minipage}
    \hfill
    \begin{minipage}{0.48\textwidth}
        \centering
        
        \begin{subfigure}[b]{\linewidth}
            \centering
            \includegraphics[width=\linewidth]{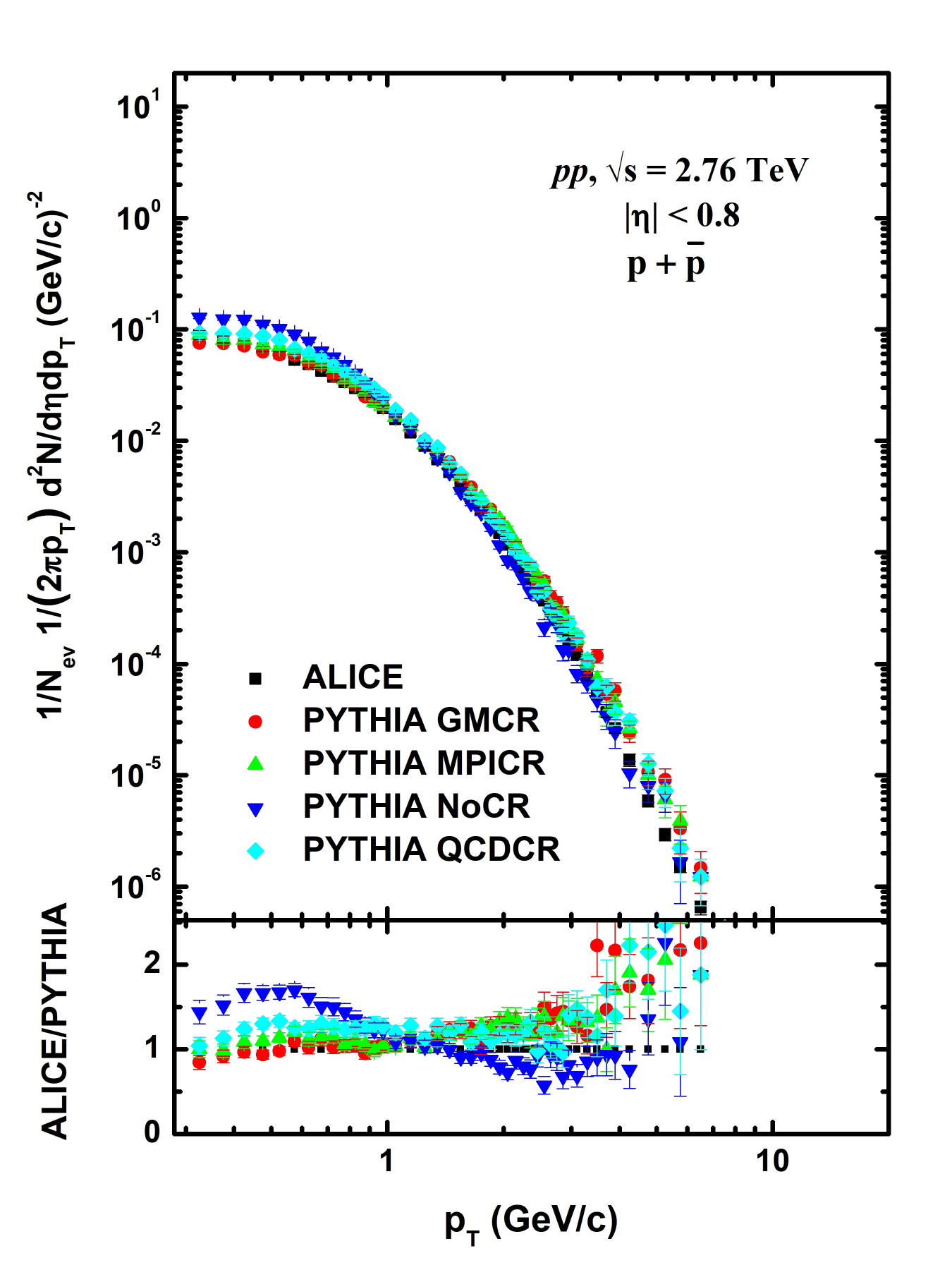}
            \caption{}
            \label{fig:1c}
        \end{subfigure}
        
        \vspace{1em}
        
            \caption{\textcolor{black}{Transverse momentum ($p_T$) spectra of (a) charged pions ($\pi^+ + \pi^-$), (b) charged kaons ($K^+ + K^-$), and (c) proton and anti-protons ($p + \bar{p}$) produced in $pp$ collisions at $\sqrt{s} = 2.76 \text{ TeV}$ for mid-rapidity ($|\eta| < 0.8$).} The ALICE data \cite{42} are compared to predictions from the PYTHIA using different CR tunes. The bottom panels show the ratio of the model predictions to the ALICE data.}
            \label{fig:1}
        
    \end{minipage}
    
\end{figure*}

We have generated $6.7 \times 10^9$ events for each individual CR tune in PYTHIA Monash for pp collisions at $\sqrt{s} =$ 2.76 TeV. To verify the datasets, we have plotted Fig.~\ref{fig:0} where the pseudorapidity distribution of primary charged particles for PYTHIA under various CR tunes is compared with ALICE experimental data \cite{32}.
From the figure, it is observed that the PYTHIA MPICR tune is in good agreement with the ALICE data, exhibiting deviations of less than $0.2\sigma$ across the full $\eta$ range. In contrast, the PYTHIA QCDCR and PYTHIA GMCR configurations show noticeable discrepancies when compared to the ALICE measurements; however, these differences remain within the associated uncertainty ranges. On the other hand, the PYTHIA NoCR setup demonstrates a significant disagreement with the ALICE data, with considerably large deviations, although it follows a similar trend.
So, to further verify the datasets, we have plotted Fig.~\ref{fig:1}, where the transverse momentum ($p_T$) spectra of charged particles are compared with ALICE experimental data \cite{42} within $|\eta|<0.8$. In Fig.~\ref{fig:1a}, the transverse momentum spectra of charged pions ($\pi^{\pm}$) are compared with ALICE data for different PYTHIA CR tunes. From the ratio plot of the figure, it is visible that PYTHIA NoCR shows a very large discrepancy around $97-100\%$ in some $p_T$ bins, meaning the model overestimates the data. Although for other setups the discrepancy is around $25-48\%$. Similarly, in Fig.~\ref{fig:1b} for charged kaons ($K^{\pm}$), PYTHIA NoCR shows a discrepancy of approximately $48\%$, which is significantly larger than for other setups. Likewise, in Fig.~\ref{fig:1c} for protons and anti-protons ($p\bar{p}$), the discrepancy for PYTHIA NoCR is also significantly larger than for other setups. Therefore, PYTHIA NoCR does not agree well with the ALICE experimental data, whereas PYTHIA with other CR tunes also shows some discrepancy, but those are either within the ALICE total uncertainties or low compared to PYTHIA NoCR.

\newgeometry{left=0.5cm, right=0.5cm, top=1cm, bottom=2cm}

\restoregeometry

\section{Results and Discussions}

In Table~\ref{tab:Table1}, we present both the uncorrected ($\nu_{(+-,dyn)}$) and corrected ($\nu_{(+-,dyn)}^{\rm corr}$) values of dynamical net-charge fluctuations from ALICE and from different PYTHIA CR tunes (NoCR, MPICR, GMCR, QCDCR) for $pp$ collisions at $\sqrt{s} = 2.76$ TeV, at $\Delta\eta$ widths of 1.0 and 1.6. The corrected values are obtained using Eq.(\ref{corrected}). As noted in Ref.~\cite{22,29}, such corrections for global charge conservation and finite acceptance effects are known to bring both experimental data and model calculations closer to Hadron Resonance Gas (HRG) expectations. A comparison between uncorrected and corrected results is therefore of significant interest to the field.
As expected, the correction shifts the values upward, making them less negative, bringing the results closer to HRG expectations. However, the relative performance of each CR tune remains the same -- PYTHIA NoCR continues to exhibit the best agreement with ALICE data, even after the correction is applied.
The absolute value of $\nu_{(+-,\text{dyn})}$ for all PYTHIA CR tuned setups, including the ALICE data, is observed to decrease as we move from a $\Delta\eta = 1.0$ to $\Delta\eta = 1.6$. While the previous analysis showed that PYTHIA NoCR significantly fails to describe the $p_T$ spectra and $\eta-$distribution (Fig. \ref{fig:0} and \ref{fig:1}), Table \ref{tab:Table1} reveals a counter-intuitive result: the $\nu_{(+-,\text{dyn})}$ value predicted by the NoCR setup shows the closest agreement with the ALICE experimental result compared to any other CR tune. A same trend is observed for the corrected values $\nu_{(+-,dyn)}^{\rm corr}$, where the NoCR configuration continues to exhibit the closest agreement with the corrected ALICE data. The PYTHIA MPICR tune with $\text{range} = 0.5$ shows better agreement with the ALICE $\nu_{(+-,\text{dyn})}$ values for both uncorrected and corrected cases. Conversely, increasing the reconnection range parameter leads to a more negative $\nu_{(+-,\text{dyn})}$ value, which directly increases the discrepancy compared to the ALICE experimental value. This behavior strongly suggests that the reconnection range is a sensitive parameter governing net-charge fluctuation.
Similarly, PYTHIA QCDCR predicts significantly higher $\nu_{(+-,\text{dyn})}$ values than ALICE data. For instance, PYTHIA GMCR exhibits the largest deviation, with a discrepancy of approximately $75\%$ from ALICE data at $\Delta\eta = 1.0$. Thus, the implementation of CR conditions in PYTHIA is essential for reproducing the single-particle spectra, but these may be over-suppressing the dynamical net-charge fluctuations in $pp$ collisions at $\sqrt{s} =$ 2.76 TeV.
\textcolor{black}{Throughout this analysis, $\langle N_{\rm ch} \rangle$ is evaluated within the same kinematic acceptance as the fluctuation observable, specifically within the pseudorapidity range under consideration ($\Delta\eta$) and transverse momentum range $0.2 < p_T < 5.0$ GeV/$c$. This ensures consistency between the fluctuation measure and the multiplicity scaling.}

\begin{figure*}[p]
    \begin{minipage}{0.5\textwidth}
        \centering
        
        \begin{subfigure}[b]{\linewidth}
            \centering
            \includegraphics[width=\linewidth]{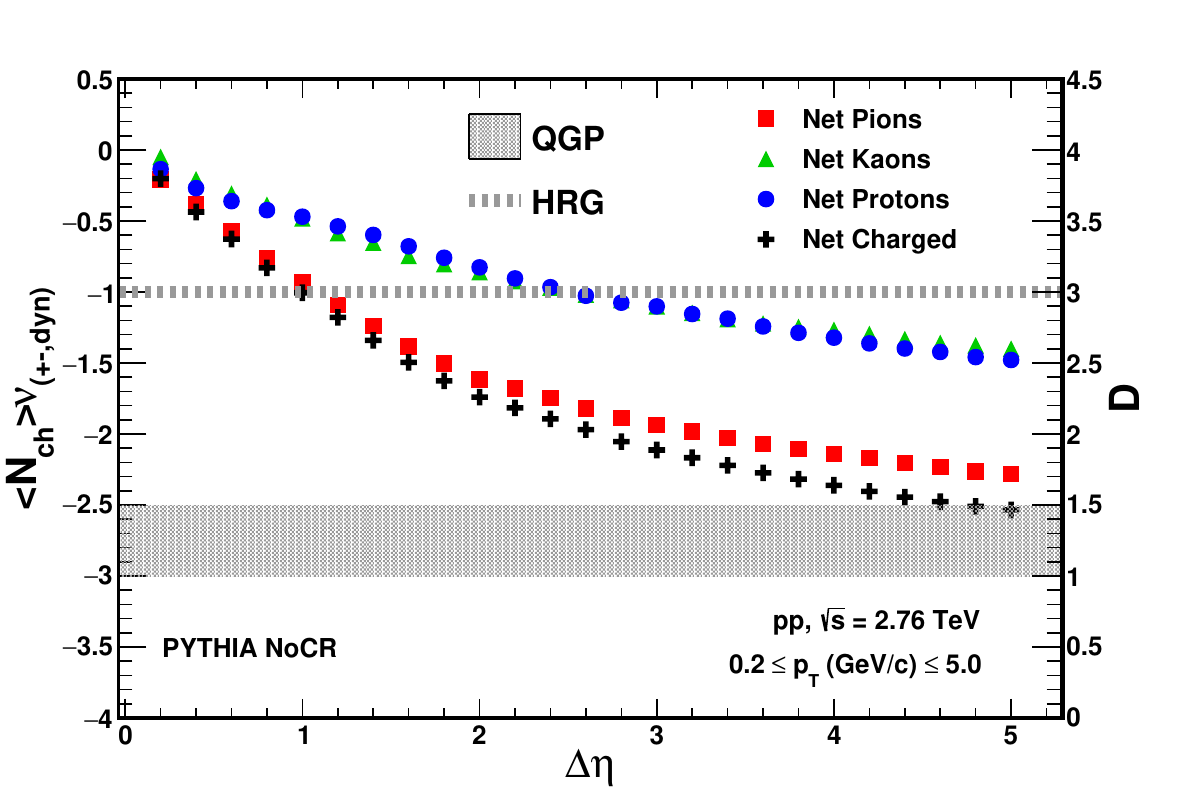}
            \caption{}
            \label{fig:a}
        \end{subfigure}

        \vspace{0.3em}

        \begin{subfigure}[b]{\linewidth}
            \centering
            \includegraphics[width=\linewidth]{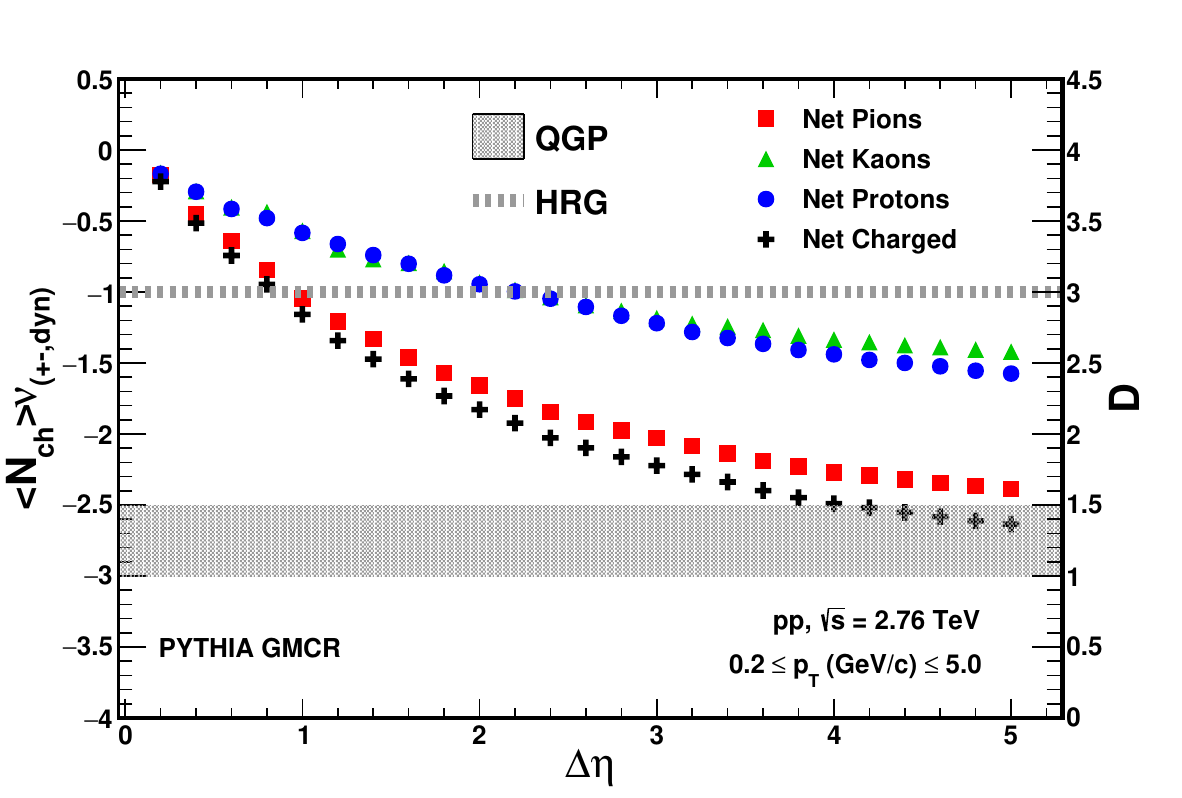}
            \caption{}
            \label{fig:b}
        \end{subfigure}

        \vspace{0.3em}

        \begin{subfigure}[b]{\linewidth}
            \centering
            \includegraphics[width=\linewidth]{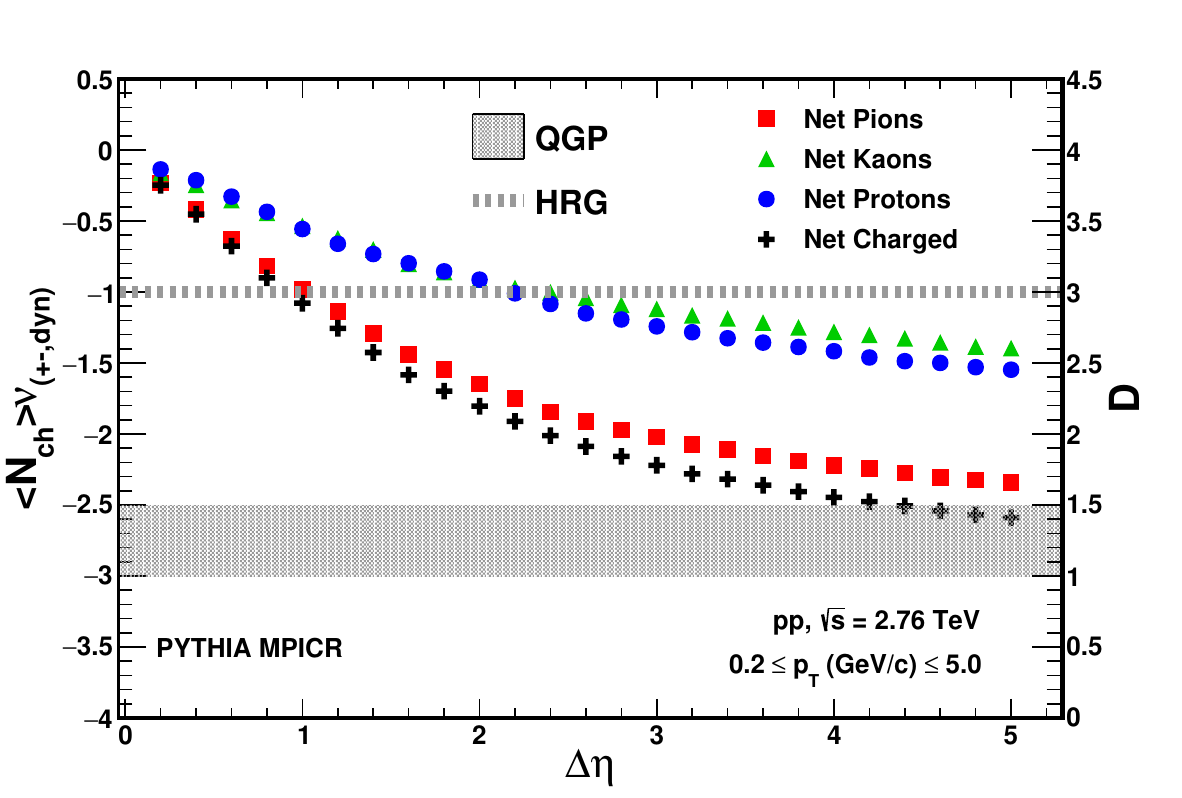}
            \caption{}
            \label{fig:c}
        \end{subfigure}
    \end{minipage}
    \hfill
    \begin{minipage}{0.5\textwidth}
        \centering
        \begin{subfigure}[b]{\linewidth}
            \centering
            \includegraphics[width=\linewidth]{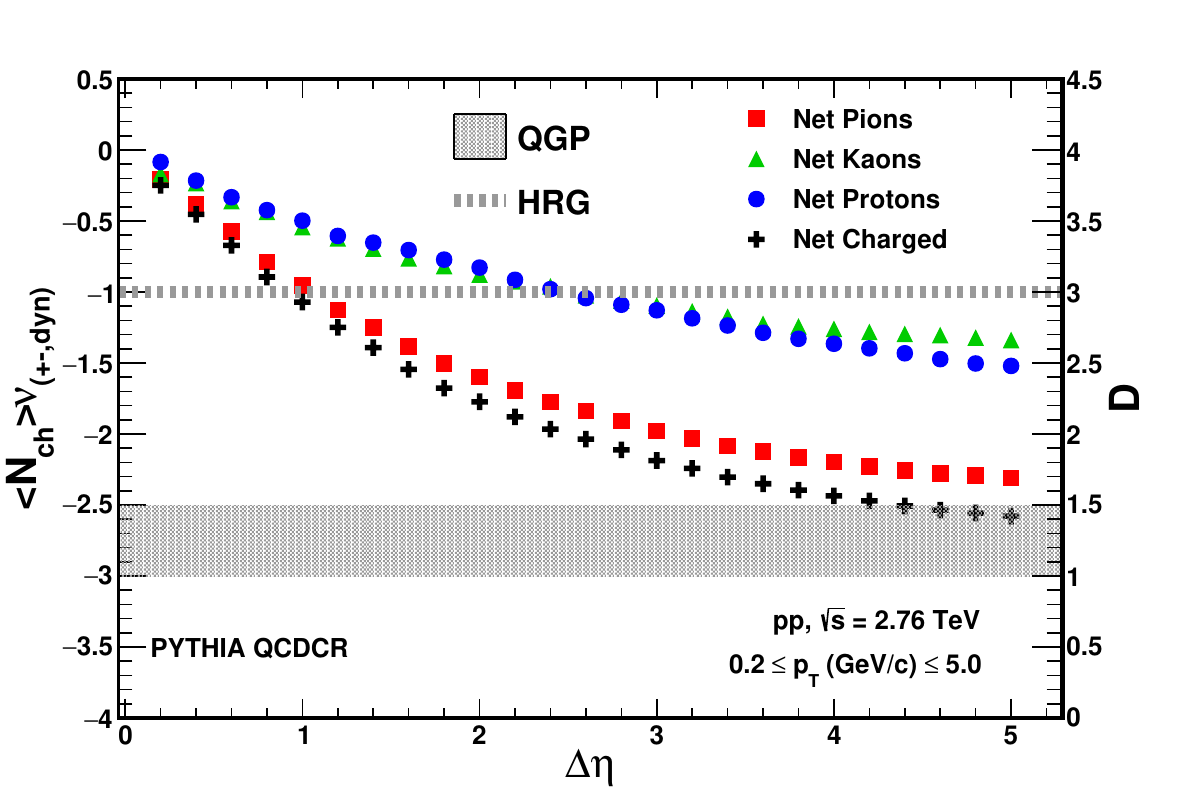}
            \caption{}
            \label{fig:d}
        \end{subfigure}
        
        \vspace{1em}
        
        \caption{\textcolor{black}{The variation of $\langle N_{\rm ch} \rangle \nu_{(+-,dyn)}$ (uncorrected for global charge conservation) as a function of the $\Delta\eta$ for net charge, net pions, net kaons, and net protons. The results are presented for PYTHIA using different CR tunes as (a) NoCR, (b) GMCR, (c) MPICR, and (d) QCDCR for $pp$ collisions at $\sqrt{s} = 2.76$ TeV. The horizontal dotted line indicates the theoretical expectation for a hadron resonance gas (HRG, $D \approx 3$) \cite{14,15,16,29}, while the band represents the QGP expectation ($D \approx 1.0$-$1.5$) \cite{14,15,16,29}.}}
        \label{fig:main}
    \end{minipage}
\end{figure*}

\begin{figure*}[p]
    \begin{minipage}{0.5\textwidth}
        \centering
        
        \begin{subfigure}[b]{\linewidth}
            \centering
            \includegraphics[width=\linewidth]{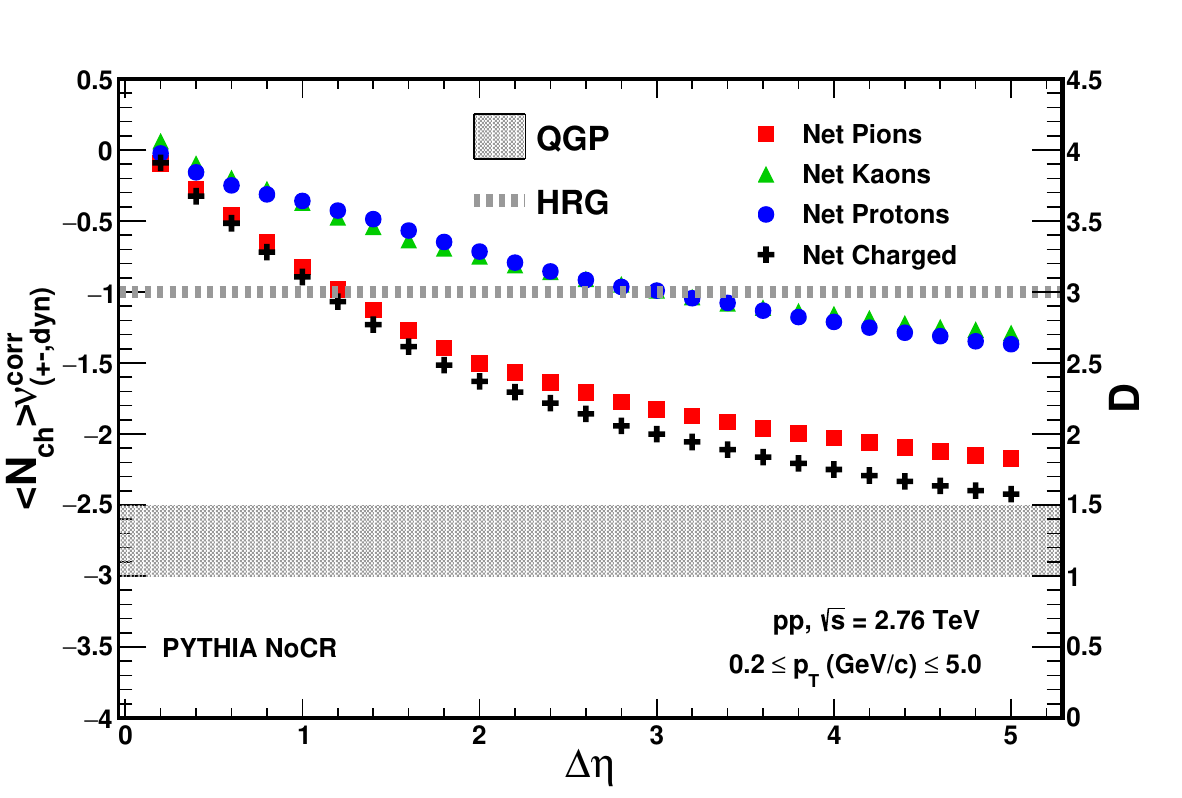}
            \caption{}
            \label{fig:a1}
        \end{subfigure}

        \vspace{0.3em}

        \begin{subfigure}[b]{\linewidth}
            \centering
            \includegraphics[width=\linewidth]{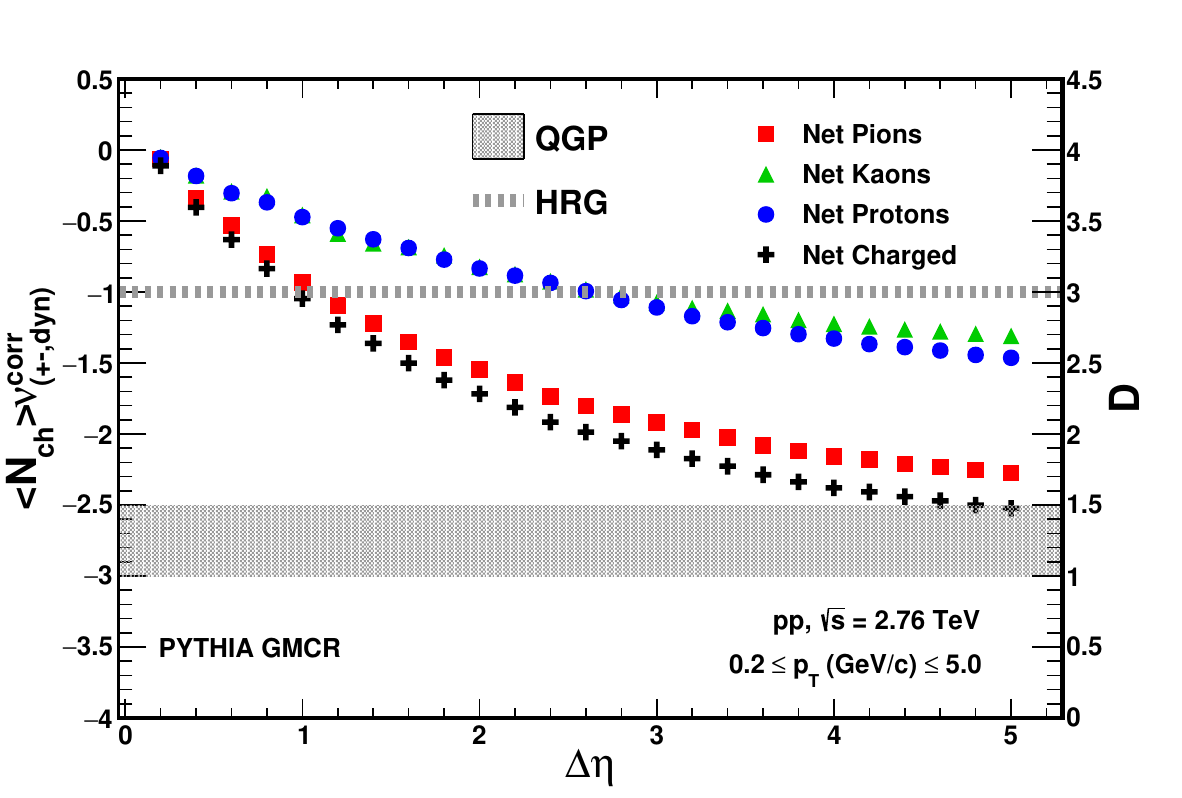}
            \caption{}
            \label{fig:b1}
        \end{subfigure}

        \vspace{0.3em}

        \begin{subfigure}[b]{\linewidth}
            \centering
            \includegraphics[width=\linewidth]{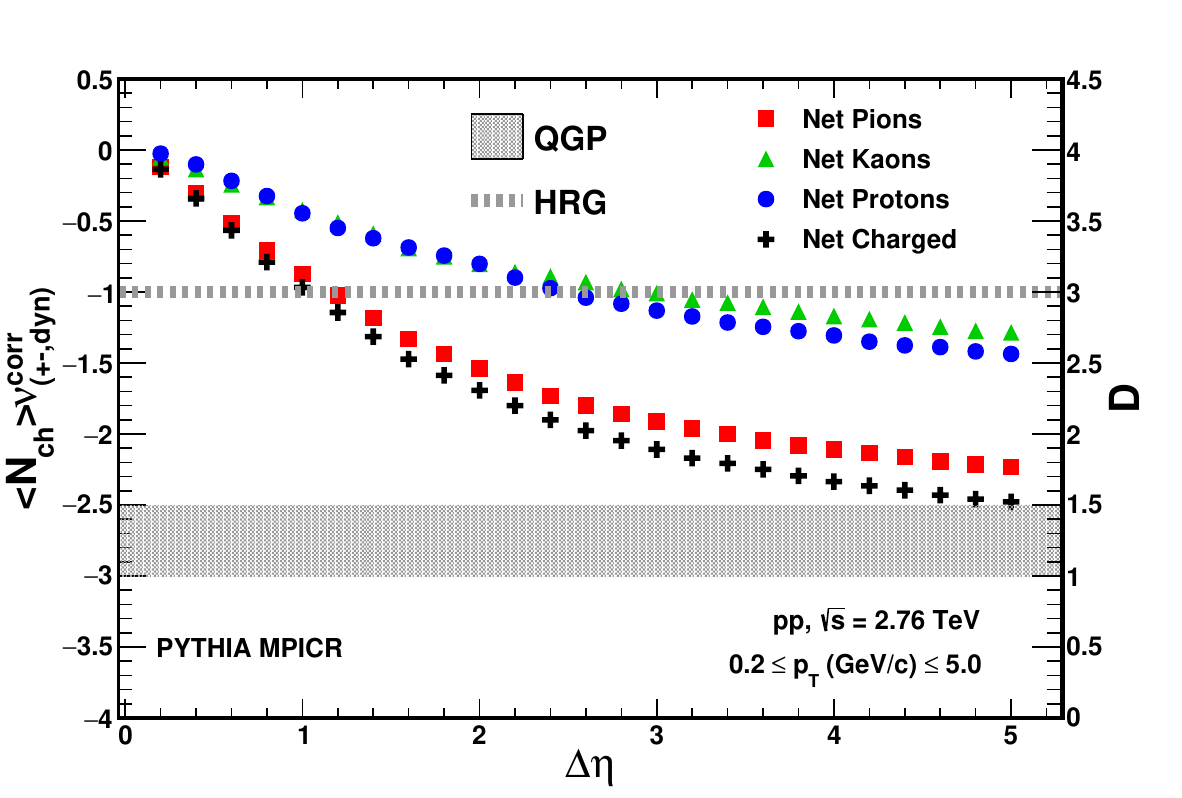}
            \caption{}
            \label{fig:c1}
        \end{subfigure}
    \end{minipage}
    \hfill
    \begin{minipage}{0.5\textwidth}
        \centering
        \begin{subfigure}[b]{\linewidth}
            \centering
            \includegraphics[width=\linewidth]{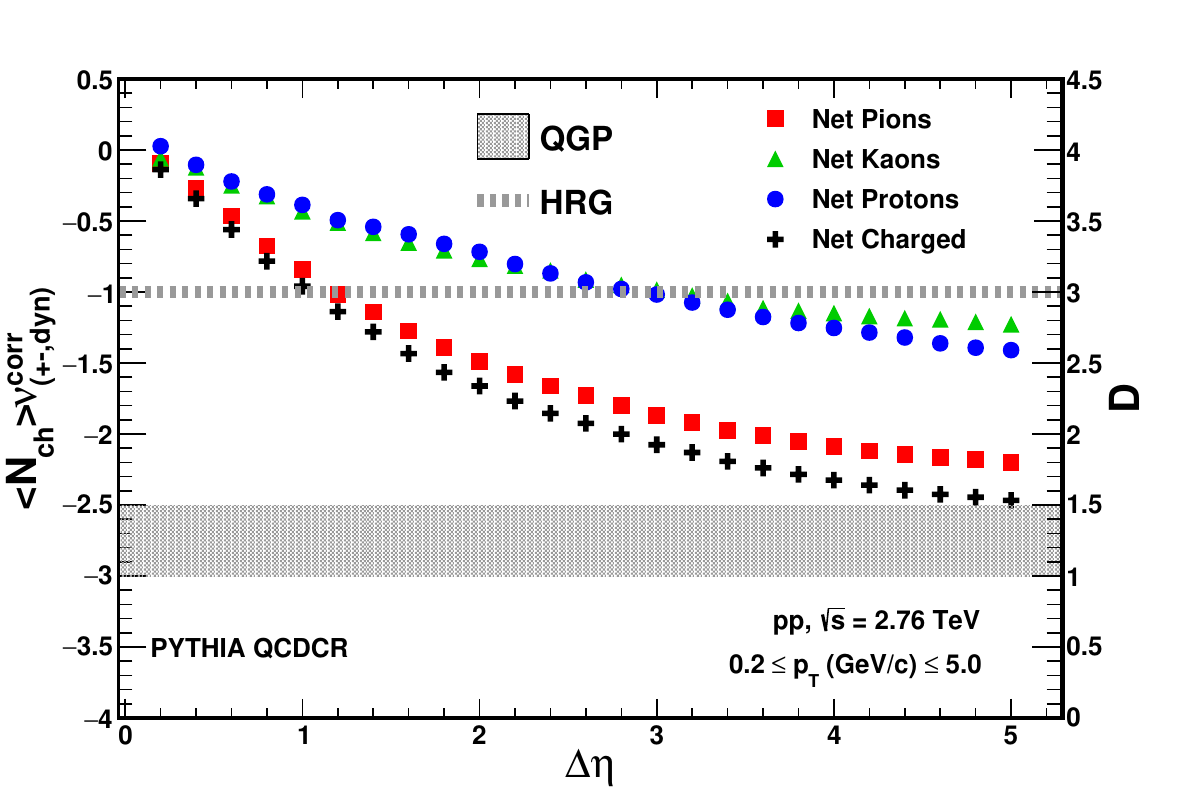}
            \caption{}
            \label{fig:d1}
        \end{subfigure}
        
        \vspace{1em}
        
        \caption{\textcolor{black}{The variation of $\langle N_{\rm ch} \rangle \nu_{(+-,dyn)}^{\rm corr}$ (corrected for global charge conservation) as a function of the $\Delta\eta$ for net charge, net pions, net kaons, and net protons. The results are presented for PYTHIA using different CR tunes as (a) NoCR, (b) GMCR, (c) MPICR, and (d) QCDCR for $pp$ collisions at $\sqrt{s} = 2.76$ TeV. The correction $ \nu_{(+-,dyn)}^{\rm corr} = \nu_{(+-,dyn)} + 4/\langle N_{\rm total}\rangle $ \cite{22,29} shifts the values upward, bringing them closer to the HRG expectations. The horizontal dotted line indicates the HRG limit ($D \approx 3$) \cite{14,15,16,29}, while the band represents the QGP expectation ($D \approx 1.0$-$1.5$) \cite{14,15,16,29}.}}
        \label{fig:main1}
    \end{minipage}
\end{figure*}

\textcolor{black}{To contextualize the magnitude of the observed fluctuations, we compare our PYTHIA results with the theoretical expectations for the hadron resonance gas (HRG) and quark-gluon plasma (QGP) phases, as indicated in Figs.~\ref{fig:main} and~\ref{fig:main1}. The net-charge fluctuations per unit entropy, quantified by the $D$ measure, are theoretically predicted to differ significantly between these two phases. In the HRG phase, where the charge carriers are hadrons with unit electric charge, the fluctuations are expected to be larger. Theoretical estimates for a simple uncorrelated pion gas yield $D = 4$, corresponding to $\langle N_{\rm ch} \rangle \nu_{(+-,dyn)}^{\rm corr} \approx 0$ \cite{15,16}. However, the inclusion of resonance decays, such as $\rho^0 \to \pi^+ + \pi^-$, reduces the net-charge fluctuations to $D \approx 3$, corresponding to $\langle N_{\rm ch} \rangle \nu_{(+-,dyn)}^{\rm corr} \approx -1$ \cite{14,15,16,29}. This convention, with HRG at $D \approx 3$, is consistent with the ALICE and CMS collaborations \cite{29,cms}. In the QGP phase, where the charge carriers are deconfined quarks with fractional charges, the fluctuations are significantly suppressed. Lattice QCD calculations and theoretical estimates yield $D \approx 1.0$ to $1.5$, corresponding to $\langle N_{\rm ch} \rangle \nu_{(+-,dyn)}^{\rm corr} \approx -2.2$ to $-2.5$ \cite{15,16,29}. This suppression arises because the fundamental charge carriers in the QGP carry smaller effective charges, leading to reduced fluctuations in the net charge per unit entropy.}

We now present our PYTHIA results for the variation of $\langle N_{\rm ch} \rangle \nu_{(+-,dyn)}$ as a function of the pseudorapidity range ($\Delta \eta$) for both net-charge and identified stable particles, using different PYTHIA CR tunes for $pp$ collisions at $\sqrt{s} = 2.76$ TeV within $0.2 < p_T < 5.0$ GeV/$c$ and $\Delta \eta \leq 5$. The uncorrected results are presented in Fig.~\ref{fig:main}. Subfigures (a), (b), (c), and (d) correspond to PYTHIA tunes NoCR, GMCR, MPICR, and QCDCR, respectively, all showing qualitatively similar behavior. For all PYTHIA tunes, the magnitude of $\langle N_{\rm ch} \rangle \nu_{(+-,dyn)}$ increases (becomes more negative) monotonically as the $\Delta \eta$ range increases. This trend indicates that the correlation between opposite-sign charged particles strengthens as the rapidity acceptance width increases, due in part to the particle production mechanism and in part to global charge conservation. For a sufficiently wide rapidity range, the net charge must add up to +2 (the total charge of the $pp$ system), which drives $\langle N_{\rm ch} \rangle \nu_{(+-,dyn)}$ toward a no-fluctuation limit. This effect is most clearly observed for pions, as they are the most abundantly produced species.

\textcolor{black}{The corresponding corrected values, obtained using Eq.~(\ref{corrected}), are presented in Fig.~\ref{fig:main1}. This dual presentation addresses the different approaches adopted in the literature: the CMS collaboration has presented net-charge fluctuation results without correcting for global charge conservation \cite{cms}, while the ALICE collaboration has shown results after applying the charge conservation correction \cite{29}. By showing both, we provide a comprehensive view of the fluctuation strength and demonstrate the effect of the global charge conservation correction on the PYTHIA predictions. A comparison between uncorrected and corrected results is therefore of significant interest to the field. As expected, the correction shifts the values upward, making them less negative and bringing the results closer to HRG expectations. However, the relative ordering among different CR tunes remains unchanged, confirming that the observed discrepancies between PYTHIA and experimental data are genuine dynamical effects rather than artifacts of global charge conservation.}

\textcolor{black}{It is important to emphasize that the approach of our PYTHIA results toward the QGP boundary at large $\Delta\eta$ ($\Delta\eta \gtrsim 4$) does not, by itself, constitute definitive evidence for QGP formation in small collision systems. Rather, this behavior reflects the combined influence of several factors: (i) the increasing importance of global charge conservation over large rapidity acceptances, which drives $\langle N_{\rm ch} \rangle \nu_{(+-,dyn)}^{\rm corr}$ toward more negative values; (ii) the specific correlation dynamics implemented in the PYTHIA color reconnection model, which actively enhances local charge correlations through the topological rearrangement of color strings; and (iii) the dominance of pions among the produced particles, which are the most abundant species and therefore have the strongest statistical weight in the inclusive net-charge measurement. These findings underscore the critical importance of performing systematic model-to-data comparisons before attributing signatures observed at large rapidity separations to collective behavior or deconfinement in small collision systems.}

\begin{table}[!ht]
\centering
\caption{Values of raw dynamical net-charge fluctuations $\nu_{(+-,dyn)}$, corrected fluctuations $\nu_{(+-,dyn)}^{\rm corr}$, and \textcolor{black}{the corresponding accepted charged-particle multiplicities $\langle N_{\text{ch}} \rangle$ from ALICE data and PYTHIA configurations} for minimum-bias $pp$ collisions at $\sqrt{s} = 2.76$ TeV. Results are presented for $\Delta\eta$ ranges of 1.0 and 1.6, with the MPICR tune results shown for different values of its reconnection range parameter. \textcolor{black}{The statistical uncertainties on $\langle N_{\rm ch} \rangle$ are negligibly small and therefore not mentioned here.}}
\label{tab:Table1}
\begin{tabular}{c c c c c c}
\toprule
Data/Model & $\Delta \eta$ & \textcolor{black}{$\langle N_{\text{ch}} \rangle$} & $\nu_{(+-,dyn)}$ & $\nu_{(+-,dyn)}^{\rm corr}$ & range \\
\midrule
ALICE Ref.\cite{29} & 1.0 & \textcolor{black}{4.8}  & -0.241 $\pm$ 0.012 & -0.130 $\pm$ 0.012 &  \\
                    & 1.6 & -- & -0.214 $\pm$ 0.012 & -0.104 $\pm$ 0.012 &  \\ 
\midrule
PYTHIA NoCR         & 1.0 & \textcolor{black}{5.2} & -0.268 $\pm$ 0.013 & -0.165 $\pm$ 0.012 &  \\
                    & 1.6 & \textcolor{black}{8.4} & -0.249 $\pm$ 0.013 & -0.147 $\pm$ 0.012 &  \\ 
\midrule
PYTHIA MPICR        & 1.0 & \textcolor{black}{4.7} & -0.276 $\pm$ 0.014 & -0.166 $\pm$ 0.012 & 0.5 \\
                    & 1.6 & \textcolor{black}{7.5} & -0.276 $\pm$ 0.014 & -0.166 $\pm$ 0.012 & \\
                    \cmidrule(lr){2-6}
                    & 1.0 & \textcolor{black}{4.8} & -0.319 $\pm$ 0.016 & -0.210 $\pm$ 0.012 & 1.0 \\
                    & 1.6 & \textcolor{black}{7.6} & -0.308 $\pm$ 0.015 & -0.199 $\pm$ 0.012 &  \\
                    \cmidrule(lr){2-6}
                    & 1.0 & \textcolor{black}{4.8} & -0.329 $\pm$ 0.016 & -0.219 $\pm$ 0.015 & 1.8 \\
                    & 1.6 & \textcolor{black}{7.7} & -0.314 $\pm$ 0.016 & -0.205 $\pm$ 0.014 & \\
                    \cmidrule(lr){2-6}
                    & 1.0 & \textcolor{black}{4.9} & -0.354 $\pm$ 0.018 & -0.244 $\pm$ 0.016 & 2.5 \\
                    & 1.6 & \textcolor{black}{7.7} & -0.344 $\pm$ 0.017 & -0.235 $\pm$ 0.016 & \\
                    \cmidrule(lr){2-6}
                    & 1.0 & \textcolor{black}{4.9} & -0.366 $\pm$ 0.018 & -0.256 $\pm$ 0.017 & 3.5 \\
                    & 1.6 & \textcolor{black}{7.9} & -0.360 $\pm$ 0.017 & -0.250 $\pm$ 0.015 & \\
                    \cmidrule(lr){2-6}
                    & 1.0 & \textcolor{black}{5.0} & -0.369 $\pm$ 0.018 & -0.260 $\pm$ 0.017 & 5.0 \\
                    & 1.6 & \textcolor{black}{7.9} & -0.360 $\pm$ 0.017 & -0.251 $\pm$ 0.014 &  \\
\midrule
PYTHIA GMCR         & 1.0 & \textcolor{black}{4.6} & -0.422 $\pm$ 0.022 & -0.304 $\pm$ 0.020 &  \\
                    & 1.6 & \textcolor{black}{7.4} & -0.371 $\pm$ 0.019 & -0.254 $\pm$ 0.017 &  \\
\midrule
PYTHIA QCDCR        & 1.0 & \textcolor{black}{4.5} & -0.369 $\pm$ 0.018 & -0.255 $\pm$ 0.018 &  \\
                    & 1.6 & \textcolor{black}{7.2} & -0.337 $\pm$ 0.017 & -0.224 $\pm$ 0.016 &  \\                        
\bottomrule     
\end{tabular}
\end{table}

\newgeometry{left=0.5cm, right=0.5cm, top=1cm, bottom=2cm}

\restoregeometry

\newgeometry{left=0.5cm, right=0.5cm, top=1cm, bottom=2cm}

\restoregeometry

Finally, the $\langle N_{\text{ch}} \rangle \nu_{(+-,\text{dyn})}$ value for net-charged particles approaches the theoretical QGP boundary predicted in Refs.~\cite{15,16,22,29} only at the highest $\Delta \eta$ values ($\Delta \eta \gtrsim 4$) for all PYTHIA tunes. For charged pions ($\pi^\pm$), the values are very close to this theoretical boundary, though they do not reach it.

\begin{figure}[!ht]
\begin{center}
\includegraphics[scale=0.52]{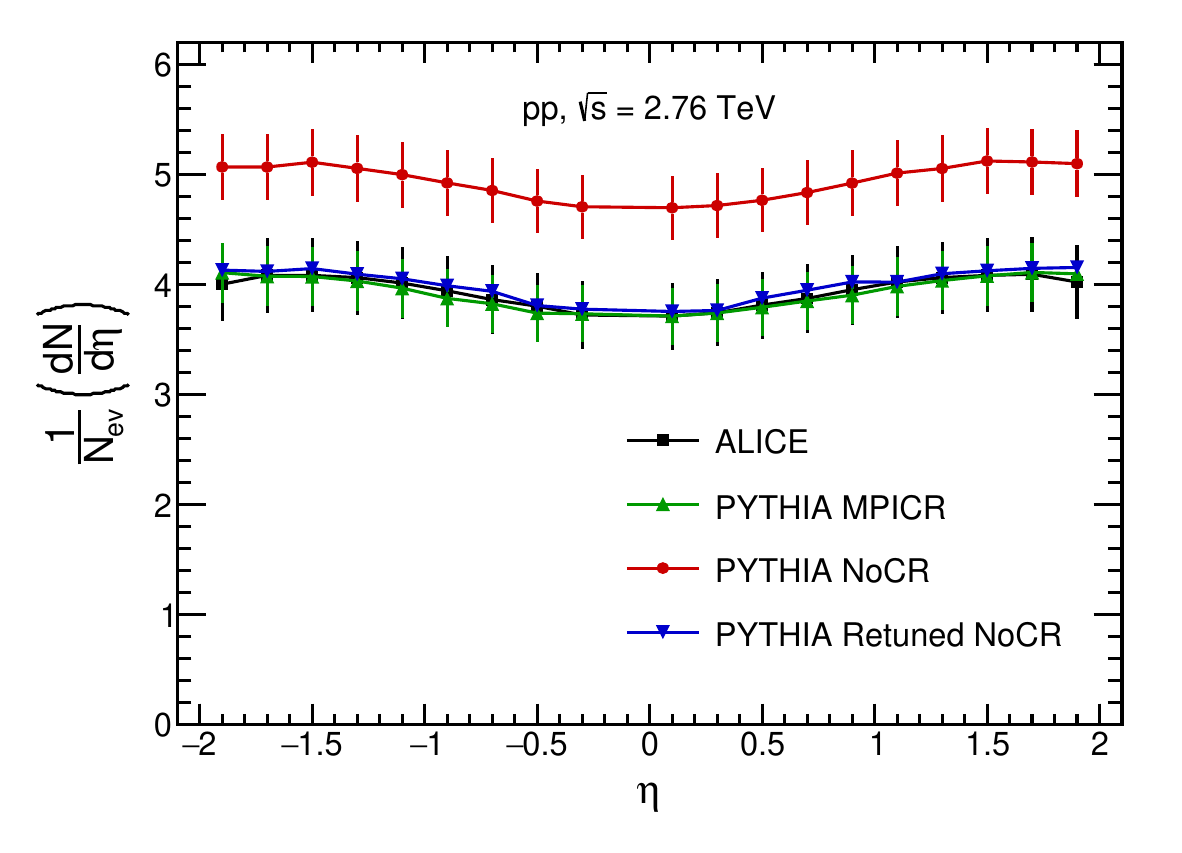} 
\caption{Pseudorapidity density ($dN/d\eta$) of primary charged particles for different PYTHIA configurations in $pp$ collisions at $\sqrt{s} = 2.76$~TeV. The default NoCR tune (red circles) shows significantly higher multiplicity compared to ALICE data (black squares) \cite{32}. After retuning the MPI parameter ($pT0Ref$), the Retuned NoCR configuration (blue inverted triangles) matches both the experimental data and the default MPICR tune (green triangles). This controlled setup isolates the dynamical effects of color reconnection from trivial multiplicity scaling.}
\label{fig:retune}
\end{center}
\end{figure}

Furthermore, it is important to distinguish whether the observed trends are driven by the specific dynamics of the color reconnection (CR) mechanism or by variations in the average charged-particle multiplicity. As illustrated in Fig.~\ref{fig:0}, disabling CR within the PYTHIA 8 framework leads to a significant increase in multiplicity, which may, in turn, influence the behavior of the fluctuation observable $\nu_{+-,dyn}$ through trivial statistical scaling. To isolate the dynamical contributions of CR from these multiplicity-dependent effects, we implemented a controlled simulation set labeled ``PYTHIA Retuned NoCR". In this configuration, CR was disabled while the multiparton interaction (MPI) parameter (``$MultipartonInteractions:pT0Ref$") was systematically increased. This adjustment ensures that the final charged-particle density ($dN/d\eta \approx 3.96$) matches both the experimental data from the ALICE Collaboration and the default 'MPICR' predictions, thereby providing a robust baseline for evaluating the impact of color reconnection at a fixed multiplicity. 
The parameter ``$pT0Ref$" in the PYTHIA framework serves as a semi-hard regularization scale for the partonic cross-section, effectively screening the $1/p_T^4$ divergence as $p_T \to 0$ \cite{39,40.2,40,40.1}. Physically, it represents the transverse momentum scale below which the proton's color charge is effectively screened. In this work, the ``$pT0Ref$" parameter was utilized as a control variable to decouple the multiplicity dependence of $\nu_{(+-,dyn)}$. By increasing ``$pT0Ref$" in the NoCR configuration, we suppressed the number of MPI-induced strings to match the experimental charged-particle density. This procedure ensures that the observed sensitivity of net-charge fluctuations to color reconnection originates from the topological rearrangement of the color fields rather than fluctuations in the total number of produced sources. The effectiveness of this matching procedure is illustrated in Fig.~\ref{fig:retune}, which compares the pseudorapidity densities of the default NoCR, Retuned NoCR, and MPICR configurations with ALICE data \cite{32}. Fig.~\ref{fig:retune} confirms that our retuning of $pT0Ref$ successfully brings the Retuned NoCR multiplicity (blue inverted triangles) into good agreement with both the ALICE data \cite{32} and the MPICR baseline. By matching the multiplicity of the NoCR setup to the data, we observe that the fluctuation value is indeed modified compared to the high-multiplicity NoCR case. However, a significant gap remains between the Retuned NoCR and the MPICR results at fixed multiplicity. Specifically, at $\Delta\eta=1.0$, the Retuned NoCR baseline is $-0.2868\pm0.0132$, while the PYTHIA MPICR (default range 1.8) shifts this to $-0.3286\pm0.0164$. This shift represents the intrinsic dynamical effect of color reconnection, proving that CR actively enhances local charge correlations independently of its effect on total particle production.

\section{Conclusions}
\label{sec:summary}

In this work, we have presented a systematic investigation into the role of Color Reconnection (CR) mechanisms in modulating net-charge fluctuations within $pp$ collisions at $\sqrt{s} = 2.76$~TeV. By comparing various PYTHIA Monash configurations-NoCR, MPICR, GMCR, and QCDCR against ALICE experimental data. The main findings and their physical implications are summarized as follows:

\begin{itemize}

	\item The comparison between uncorrected and corrected $\nu_{(+-,dyn)}$ values in Table~\ref{tab:Table1} shows that while the global charge conservation correction shifts the magnitude upward as expected, the relative ordering among different CR tunes remains unchanged. 

    \item  While the MPICR tune provides the most accurate description of the charged-particle pseudorapidity density ($\text{d}N/\text{d}\eta$) and transverse momentum ($p_T$) spectra, the NoCR configuration despite significantly overestimating the particle yields produces $\nu_{(+,-,\text{dyn})}$ values in closer quantitative agreement with experimental observations. This suggests a phenomenological inconsistency in the current implementation of CR within the Monash tune: while CR is essential for describing the energy flow via string-length minimization, it appears to simultaneously introduce an over-correlation of local charge partners.
    
    \item  To decouple the trivial influence of multiplicity from genuine dynamical correlations, we utilized a ``Retuned NoCR'' setup. By adjusting the semi-hard regularization scale, ``$pT0Ref$", which represents the transverse momentum scale below which the proton's color charge is effectively screened \cite{2,40,40.1}, we matched the charged-particle density of the NoCR case to the ALICE baseline. The persistence of a significant gap between the Retuned NoCR baseline ($\nu_{(+,-,\text{dyn})} \approx -0.2868 \pm 0.0132$) and the MPICR results ($\nu_{(+,-,\text{dyn})} \approx -0.3286 \pm 0.0164$) at identical multiplicities proves that CR acts as a topological rearrangement of the color fields rather than a simple statistical scaling of the particle density.
    
    \item Our analysis revealed that $\nu_{(+,-,\text{dyn})}$ exhibits a strong sensitivity to the MPICR reconnection range parameter. Increasing the range leads to more negative values, indicating that the geometric extent over which color strings are allowed to interact is a primary driver of the strength of opposite-sign particle correlations.
    
    \item The investigation of identified particle channels reveals that the approach toward the model-predicted QGP boundary at large $\Delta\eta$ is predominantly confined to the inclusive net-charge and net-pion sectors. The absence of similar behavior in the kaon and proton channels indicates that these correlations are sensitive to the underlying production dynamics of strange quarks, as well as to the more intricate recombination processes involved in baryon formation.
    
    \item \textcolor{black}{By presenting both uncorrected (Fig.~\ref{fig:main}) and corrected (Fig.~\ref{fig:main1}) values of $\langle N_{\rm ch} \rangle \nu_{(+-,dyn)}$, our study bridges the two approaches adopted by the CMS \cite{cms} and ALICE \cite{29} collaborations, respectively. This dual presentation demonstrates that while the correction shifts the values upward as expected \cite{22,29,cms}, the relative ordering among different CR tunes remains unchanged, confirming that the observed discrepancies between PYTHIA and experimental data are genuine dynamical effects rather than artifacts of global charge conservation.}
    
\end{itemize}

In conclusion, net-charge fluctuations serve as a high-precision diagnostic tool for the internal mechanics of string fragmentation models. Our results highlight the need for a more nuanced balance in CR models between achieving string-length minimization and preserving the observed balance of local charge conservation. These findings provide a necessary baseline for interpreting potential collective signals in high-multiplicity $pp$ events and offer quantitative benchmarks for the future development of QCD-inspired event generators.

\section*{Data availability}

Data will be made available on request.

\section*{Declaration of competing interest}
The authors declare that they have no known competing financial
interests or personal relationships that could have appeared to influence
the work reported in this paper.

\section*{Acknowledgement}
P. K. Haldar and D. Dhar gratefully acknowledge the financial support provided by the Department of Science and Technology and Biotechnology, Government of West Bengal, India (Sanction Memo No. 1197(Sanc.)/STBT-11012(26)/4/2024-ST SEC). Also, T. Biswas acknowledges the financial support as an Inspire fellow (No. DST/INSPIRE Fellowship/2022/IF220173) from the Department of Science and Technology, Government of India.


\end{document}